\documentclass[conference]{IEEEtran}
\IEEEoverridecommandlockouts
\usepackage{cite}
\usepackage{amsmath,amssymb,amsfonts}
\usepackage[ruled,vlined,linesnumbered]{algorithm2e}
\usepackage{graphicx}
\usepackage{textcomp}
\usepackage{xcolor}
\usepackage{booktabs} 
\usepackage{multirow}
\usepackage{hyperref}
\usepackage{pifont}
\usepackage{array}
\usepackage{orcidlink}

\def\BibTeX{{\rm B\kern-.05em{\sc i\kern-.025em b}\kern-.08em
    T\kern-.1667em\lower.7ex\hbox{E}\kern-.125emX}}
\begin{document}
%%%%%%%%%%%%%%for comments%%%%%%%%
\newcommand\JP[1]{\textcolor{magenta}{#1}}
\newcommand\TP[1]{\textcolor{green}{#1}}
%%%%%%%%%%%%%%%%%%%%%%%%%%%%%%%%%%%

\title{FlowGRN+: Improving Gene Regulatory Network Inference by Spline Fitting and Manifold Projection in Conditional Flow Matching (Technical Report)
\thanks{Authors Tsz~Pan Tong and Jun Pang acknowledge financial support of the Institute for Advanced Studies (IAS) of the University of Luxembourg through an Audacity Grant (AUDACITY-2021).
    This work was also supported by the Luxembourg National Research Fund (FNR) under the grant agreement INTER/NCN/24/18732364/EdgeCR.}
}

\author{\IEEEauthorblockN{Tsz~Pan Tong\,\orcidlink{0000-0001-8111-5886}}
\IEEEauthorblockA{
Faculty of Science, Technology and Medicine \\
Institute for Advanced Studies \\
University of Luxembourg \\
%4365 
Esch-sur-Alzette, Luxembourg \\
tszpan.tong@uni.lu
}

\and
\IEEEauthorblockN{Jun Pang\,\orcidlink{0000-0002-4521-4112}}
\IEEEauthorblockA{
Faculty of Science, Technology and Medicine \\
Institute for Advanced Studies \\
University of Luxembourg \\
%4365 
Esch-sur-Alzette, Luxembourg \\
jun.pang@uni.lu
}
}

\maketitle

\begin{abstract}
Gene regulatory networks (GRNs) are fundamental in understanding cellular dynamics and underlying mechanisms during development and disease.
Although scRNA-seq technologies have enabled the collection of vast numbers of gene expression profiles at single-cell resolution, inferring GRNs from scRNA-seq data remains a significant challenge due to high dimensionality and dropout.
Recently, FlowGRN has shown promising results in reconstructing cell trajectories and inferring GRNs by applying conditional flow matching (CFM) to learn the cell dynamics.
However, FlowGRN still faces limitations in the temporal coherence of reconstructed dynamics and relies on human inspection, which hinders downstream applications and reproducibility.
In this paper, we propose FlowGRN+, an improved version of FlowGRN that integrates spline fitting into the CFM framework to generate more stable reference trajectories for training, thereby improving the temporal coherence of the learned dynamics.
To address overshooting in spline fitting, we further introduce a projection scheme that projects spline tangents onto the local tangent space of the data manifold.
We evaluate FlowGRN+ on the BEELINE benchmark and show improved trajectory smoothness with a competitive GRN inference performance.
FlowGRN+ provides a practical framework for reconstructing cell trajectories and inferring GRNs from scRNA-seq data, and the insights from this work may also be useful for other CFM-based models of cellular dynamics.
\end{abstract}


\section{Introduction}
\label{sec:introduction}
Gene regulatory networks (GRNs) are fundamental to understanding how cell states change during development and disease, yet inferring the underlying network dynamics from single-cell snapshot data remains a significant challenge.
Current droplet-based single-cell RNA sequencing (scRNA-seq) technologies cannot track the same cell's gene expression profile over time, and cells are destroyed during the assay.
The observable gene expression profiles are thus snapshots of the underlying dynamic process, making it difficult to infer the hidden GRN dynamics within the complex biological system.

Classical GRN inference methods, such as GENIE3~\cite{huynh2010genie3}, TIGRESS~\cite{haury2012tigress}, and PIDC~\cite{chan2017pidc}, discard temporal information and treat the scRNA-seq data as a static snapshot, which limits their ability to capture the dynamic nature of GRNs.
Although methods based on ordinary differential equations (ODEs), such as SCODE~\cite{matsumoto2017scode} and GRISLI~\cite{aubin2020grisli}, attempt to model cell dynamics explicitly, they suffer from oversimplification and scalability problems because cell dynamics are high-dimensional, complex, and non-linear.
Recently, deep learning frameworks such as flow matching and diffusion models enable the learning of the dynamics underlying distributional shifts in high-dimensional data, such as images and texts~\cite{ho2020denoising,austin2021structured,lipman2022flow}.
This provides a promising route for reconstructing cell dynamics and GRN inference from snapshot data.

FlowGRN~\cite{tong2025flowgrn} is the first GRN inference model that implements this idea.
Starting from observed gene expression snapshots at discrete time points, FlowGRN computes optimal transport (OT) couplings between consecutive time points to generate reference velocity vectors from linearly interpolated trajectories and trains a neural network to match the references using conditional flow matching (CFM).
The cell trajectories are reconstructed by integrating the learned vector field starting from each observed cell, and the GRN is inferred by applying dynGENIE3~\cite{huynh2018dyngenie3} on the reconstructed trajectories.
While it shows top-tier performance on both simulated and real-world datasets in the BEELINE benchmark~\cite{pratapa2020benchmarking}, it still has limitations in the temporal coherence of the learned dynamics and the need for human inspection, hindering the downstream use of trajectories and reproducibility.
First, the linear-interpolated reference trajectories introduce discontinuities in the learned vector field at the boundaries of the time interval, which is biologically unrealistic and inconsistent with the continuous nature of cell dynamics.
Moreover, linearly interpolated references discard information beyond adjacent time points, focusing only on local dynamics and potentially missing long-range temporal structure.
In addition, although FlowGRN uses a dropout-robust cost when computing OT couplings, the linearly interpolated references are also affected by dropouts, which induce sudden changes in the reference velocity vectors.
Finally, FlowGRN relies on well-separated snapshot distributions at each time point, which are not always available in real-world datasets, especially when time points are close and developmental speed is non-uniform.

While we acknowledge that CFM is a promising framework for modeling cell dynamics, we argue that the temporal coherence of the learned dynamics can be further improved by designing better reference trajectories for CFM training.
In this paper, we build on the multi-marginal flow matching (MMFM) framework~\cite{rohbeck2025modeling,lee2025multi} to propose FlowGRN+, an improved version of FlowGRN that improves the temporal coherence of the learned dynamics.
We experimentally show that smoothing splines can generate dropout-robust reference trajectories, which in turn improve the temporal coherence of the dynamics learned by CFM.
To address overshooting in spline fitting, we further introduce a projection scheme to refine the reference velocity vectors by projecting them onto the local tangent space of the data manifold.
We evaluate FlowGRN+ on the BEELINE benchmark and show improved trajectory smoothness with competitive GRN inference performance.
Our method is publicly available at \url{https://github.com/1250326/FlowGRN_plus}.

In short, our contributions are three-fold:
\begin{itemize}
    \item We propose FlowGRN+, an improved version of FlowGRN that uses spline fitting to generate dropout-robust reference trajectories for CFM training, thereby improving the temporal coherence of the learned dynamics.
    \item We introduce a projection scheme to refine the reference velocity vectors by projecting them onto the local tangent space of the data manifold, which helps mitigate overshooting in spline fitting.
    \item We evaluate FlowGRN+ on the experimental BEELINE datasets and demonstrate a trade-off: FlowGRN+ improves trajectory smoothness and achieves top-tier EPR, whereas its AUPRC drops slightly and varies compared to FlowGRN.
\end{itemize}

\section{Related Work}
\label{sec:related_work}

\subsection{Gene regulatory network inference}
\label{ssec:grn_inference}
GRN inference is a long-standing problem in computational biology and has been extensively studied in the past decades.
Although single-cell sequencing technologies enrich the available data for GRN inference, the problem remains challenging due to the high dimensionality, noise, and sparsity of the data, as well as the complex and dynamic nature of GRNs.
Correlation-based methods (ppcor~\cite{kim2015ppcor}, LEAP~\cite{specht2017leap}), information theory-based methods (ARACNe~\cite{margolin2006aracne}, CLR~\cite{faith2007clr}, PIDC~\cite{chan2017pidc}, Scribe~\cite{qiu2020scribe}) and regression methods (TIGRESS~\cite{haury2012tigress}, SINCERITIES~\cite{papili2018sincerities}, GRNVBEM~\cite{sanchez2018grnvbem}) provide a statistically sound approach to infer GRNs from limited data.
However, these methods are bound by strong statistical assumptions and less flexible in modeling complex cell dynamics.
ODE-based methods (SCODE~\cite{matsumoto2017scode}, GRISLI~\cite{aubin2020grisli}) attempt to model cell dynamics with linear ODEs, which oversimplify system complexity and are computationally inefficient on large GRNs.

Taking advantage of the rapid development of machine learning, ensemble methods (GENIE3~\cite{huynh2010genie3}, dynGENIE3~\cite{huynh2018dyngenie3}, GRNBOOST2~\cite{moerman2019grnboost2}) offer powerful and scalable options for GRN inference.
GENIE3 and GRNBOOST2 are the most established and widely used GRN inference methods, and the latter is the default option in the SCENIC~\cite{aibar2017scenic} and SCENIC+~\cite{bravo2023scenic+} protocols for coarse-grained regulatory pair screening.
Both methods train a tree ensemble model to predict the expression of each target gene from other genes and use the importance score in each model to quantify the regulatory relationships.
Jump3~\cite{huynh2015jump3} and dynGENIE3 aim to incorporate temporal information, but their reliance on continuous cellular trajectories, which are unobservable in destructive snapshot-based scRNA-seq, has largely limited their application to simulated datasets rather than real-world scenarios.
Recently, researchers have tried to borrow concepts from deep learning, such as convolutional neural networks (DeepDRIM~\cite{chen2021deepdrim}, 3DCEMA~\cite{fan2021gene}, DELAY~\cite{reagor2023delay}), variational autoencoder (DeepSEM~\cite{shu2021deepsem}), and explainable AI (LRP~\cite{keyl2023lrp}).
However, many of these emerging deep learning methods have not yet matched ensemble methods in terms of precision and scalability, and thus their application is not yet widespread.

\subsection{Conditional flow matching in cellular dynamics}
\label{ssec:cfm_cellular_dynamics}
Conditional flow matching~(CFM)~\cite{lipman2022flow} is a type of generative model for learning a vector field that pushes the source distribution toward the target.
Since scRNA-seq data are intrinsically a series of snapshot distributions of the underlying cell dynamics, CFM is a natural choice for modeling cell dynamics.
OTCFM~\cite{tong2023improving} and [SF]\textsuperscript{2}M~\cite{tong2024simulation} are the earliest simulation-free CFM approaches applied to cell trajectory reconstruction.
They use the OT coupling between cell distributions to generate a reference velocity for CFM training and avoid costly integration in neural ODE training, successfully scaling up to thousands of genes.
Following their success, metric flow matching (MFM)~\cite{kapusniak2024metric} introduces a metric learning module to capture the intrinsic geometry of the data manifold; 
Curly-FM~\cite{petrovic2025curly} extends CFM to learn non-gradient dynamics and integrate RNA velocity prior; 
and VGFM~\cite{wang2025joint} jointly considers the state transition and cell growth dynamics.
These methods highlight CFM's flexibility in modeling complex cell dynamics and integrating various biological priors.

In particular, we focus on a special type of CFM with multi-marginal coupling.
Standard CFM methods only consider couplings between two distributions, whereas scRNA-seq data typically form a sequence of distributions across multiple time points.
To leverage the temporal dependencies across multiple distributions, multi-marginal flow matching (MMFM)~\cite{rohbeck2025modeling} samples cells from OT couplings across multiple time points and fits cubic splines as reference trajectories for CFM training.
Similarly, MMSFM~\cite{lee2025multi} uses Hermite splines to fit the reference trajectories under irregular time intervals and regularizes the CFM model with score matching.
Spline fitting, together with OT coupling, provides a direct way to link a sequence of distributions and preserve long-range temporal structure, which is useful for learning more temporally coherent dynamics and reconstructing trajectories better aligned with the local manifold structure.

Table~\ref{tab:reference_path_comparison} summarizes the methodological
differences in reference-path construction among closely related CFM models in terms of coupling of marginal distributions, choice of reference paths, and data manifold handling.

\begin{table}[!t]
\caption{Comparison of conditional reference construction.\\ $\mathcal{M}$ denotes the data manifold.}
\label{tab:reference_path_comparison}
\centering
\begin{tabular}{
    >{\raggedright\arraybackslash}p{0.18\linewidth}%
    >{\raggedright\arraybackslash}p{0.10\linewidth}%
    >{\raggedright\arraybackslash}p{0.20\linewidth}%
    >{\raggedright\arraybackslash}p{0.13\linewidth}%
    >{\raggedright\arraybackslash}p{0.15\linewidth}%
}
\toprule[1pt]\midrule[0.3pt]
Method
& Marginals
& Coupling
& Ref. path
& Manifold handling \\
\midrule
OTCFM~\cite{tong2023improving}
& 2
& Pairwise OT
& Linear
& Euclidean \\\addlinespace[1.5pt]

MMFM~\cite{rohbeck2025modeling}
& Multi.
& OT chain
& Cubic spline
& Euclidean \\\addlinespace[1.5pt]

MMSFM~\cite{lee2025multi}
& Multi.
& OT chain
& Monotone Hermite
& Euclidean \\\addlinespace[1.5pt]

MFM~\cite{kapusniak2024metric}
& 2
& Pairwise OT
& Approxi. geodesic
& $\mathcal{M}$-aware metric \\\addlinespace[1.5pt]

FlowGRN~\cite{tong2025flowgrn}
& 2
& Dropout-aware pairwise OT
& Linear
& Euclidean \\\addlinespace[1.5pt]

FlowGRN+
& Multi.
& Dropout-aware OT chain
& Smooth-ing spline
& $\mathcal{M}$-aware projection \\
\midrule[0.3pt]\bottomrule[1pt]
\end{tabular}
\end{table}

\section{Preliminaries}
\label{sec:preliminaries}

\subsection{Notations and problem definition}
\label{ssec:notations}
A scRNA-seq dataset contains a series of $n$ gene expression profiles $X=\{X^{(t_0)},X^{(t_1)},\cdots,X^{(t_{n-1})}\}$, each profile $X^{(t_i)}\in\mathbb{R}^{g\times c_i}$ contains $g$ genes and $c_i$ cells sampled at time $t_i$.
A cell $x\in\mathbb{R}^{g}$ can be viewed as a point in the $g$-dimensional gene space.
Because of non-uniform developmental speeds, cells collected at different time points may overlap in the gene expression space, and thus, pseudotime is often used to represent the developmental progress of each cell.
Denote the normalized pseudotime of cell $x$ as $\tau(x)\in [0,1]$, which can be estimated by existing methods such as Monocle~\cite{van2020monocle3}, Slingshot~\cite{street2018slingshot}, and DPT~\cite{haghverdi2016diffusion}.
A GRN is a directed graph $\mathcal{G}=(\mathcal{V},\mathcal{E},\omega)$, where $\mathcal{V}$ is the set of genes, $\mathcal{E}\subset \mathcal{V}\times\mathcal{V}$ is the set of edges, and $\omega:\mathcal{E}\rightarrow\mathbb{R}$ is the weight function (regulation strength).
An adjacency matrix $A\in\mathbb{R}^{g\times g}$ is a matrix representation of the GRN, where $A_{i,j}=\omega(i,j)$ if $(i,j)\in\mathcal{E}$ and $A_{i,j}=0$ otherwise.
We aim to infer the directed GRN $\mathcal{G}$ from the scRNA-seq dataset $X$ without any prior knowledge of the GRN structure.

\subsection{Conditional flow matching}
\label{ssec:CFM}
Cellular trajectories are crucial for inferring GRNs.
In a related line of research, structural inference studies~\cite{wang2022isidg,wang2023rcsi,wang2024benchmarking} have shown that trajectory data alone can recover the underlying system dynamics.
This motivates us to reconstruct the cellular trajectories to facilitate GRN inference.

Neural ODE is a powerful framework for modeling dynamical systems: it learns a vector field whose ODE solution matches observed time-series data.
Recent methods, such as TrajectoryNet~\cite{tong2020trajectorynet} and TIGON~\cite{sha2024tigon}, have successfully applied neural ODEs to scRNA-seq data.
However, their reliance on ODE integration makes them computationally expensive and numerically unstable, limiting their applications to small gene sets ($<50$) or low-dimensional latent spaces.
To overcome these limitations, we adopted the OTCFM model~\cite{tong2023improving}, an integration-free conditional flow matching model.
OTCFM leverages optimal transport (OT) plans between snapshots to define target vector fields, thereby guiding the learning of the dynamics that drive cellular transitions.

CFM trains the vector field with respect to the displacement between the two distributions, whereas neural ODE trains the vector field with respect to the observed trajectories, thereby incurring costly ODE integration.
Viewing scRNA-seq snapshots as a sequence of gene expression distributions, CFM is a natural choice for learning the vector field that drives those flows.
CFM is a generative model that learns a vector field $v_\theta: \mathbb{R}^{g}\times [0,1]\rightarrow\mathbb{R}^{g}$ such that $\frac{dx}{dt}=v_\theta(x,t)$ pushes the source distribution $p_0$ to the target distribution $p_1$.
Given a conditioning variable $z\sim q(z)$ and a probability path $p(t|z)$ connecting $p_0$ and $p_1$, we can compute velocity vectors $u(x,t|z)$ from $p(t|z)$ to be matched with $v_\theta$ by minimizing the following objective by taking the expectation on $t\sim\mathcal{U}(0,1), z\sim q(z), x\sim p(t|z)$:
\begin{equation}
    \mathcal{L}_{CFM}(\theta) = \mathbb{E} \| v_\theta(x,t) - u(x,t|z)\|^2. \label{eq:loss_cfm}
\end{equation}

OTCFM sets $q(z)=\pi(x_0,x_1)$ as the optimal transport coupling $\pi$ between $x_0\sim p_0$ and $x_1\sim p_1$.
Then, under linear interpolation $\mu(t|x_0,x_1)$ and noise scale $\sigma$:
\begin{align}
    \mu(t|x_0,x_1) &= t x_1 + (1-t)x_0, \, \sigma(t) = \sigma\sqrt{t(1-t)}, \label{eq:cfm_mu_sigma}\\
    p(t|x_0,x_1) &= \mathcal{N}(\mu(t|x_0,x_1),\sigma(t)^2I), \label{eq:cfm_p}\\
    u(x,t|x_0,x_1) &= \frac{\sigma'(t)}{\sigma(t)}(x-\mu(t))+\mu'(t) \label{eq:cfm_u}.
\end{align}
These settings make the CFM objective computable.

\subsection{Multi-marginal flow matching}
\label{ssec:mmfm}
Multi-marginal flow matching (MMFM)~\cite{rohbeck2025modeling} extends CFM to learn the vector field that pushes a sequence of distributions $p_0,p_1,\cdots,p_{n-1}$.
Suppose each distribution $p_i$ is observed at time $t_i$, and the OT coupling between consecutive distributions $p_i$ and $p_{i+1}$ is $\pi_i$.
MMFM assumes that $q(z)$ is Markovian and can be factorized as $q(z)=q(x_0,x_1,\cdots,x_{n-1})=\prod_{i=0}^{n-2}\pi_i(x_i,x_{i+1})$.
Then, MMFM fits a cubic spline $\gamma(t)=\mu(t|x_0,x_1,\cdots,x_{n-1})$ as the reference trajectory for CFM training.
In this setting, Eqs.~\eqref{eq:cfm_p}--\eqref{eq:cfm_u} are piecewise defined in each time interval $[t_i,t_{i+1}]$ with $\mu(t|z), \sigma(t)$:
\begin{align}
    \mu(t|z) &= \gamma(t), \quad \sigma(t) = \sigma\frac{\sqrt{(t_{i+1}-t)(t-t_i)}}{t_{i+1}-t_i}, \label{eq:mmfm_mu_sigma}
\end{align}

\subsection{dynGENIE3}
\label{ssec:dynGENIE3}
dynGENIE3~\cite{huynh2018dyngenie3} is a GRN inference method that extends GENIE3~\cite{huynh2010genie3} to time-series data.
dynGENIE3 trains a random forest $f_j$ to model each target gene $j$ in cell trajectories $\{x(t)\}$ using $\frac{dx_j(t)}{dt}=f_j(x(t))-\alpha_jx_j(t)$, where $\alpha_j$ is a decay rate of gene $j$ and $\frac{dx_j(t)}{dt}$ is approximated by finite differences $\frac{x_j(t+\Delta t)-x_j(t)}{\Delta t}$.
The importance score $\mathcal{I}_j(i)$ of each random forest model $f_j$ is used to quantify the regulatory strength of the gene $i$ on $j$ and to build the adjacency matrix $A_{i, j}=\mathcal{I}_j(i)$.
This will yield a directed graph, as the importance scores $\mathcal{I}_j(i)$ and $\mathcal{I}_i(j)$ are generally unequal.

\section{Method}
\label{sec:method}

\begin{figure*}[!t]
  \centering
  \includegraphics[scale=0.85,trim=1cm 0cm 0.5cm 0cm,clip]{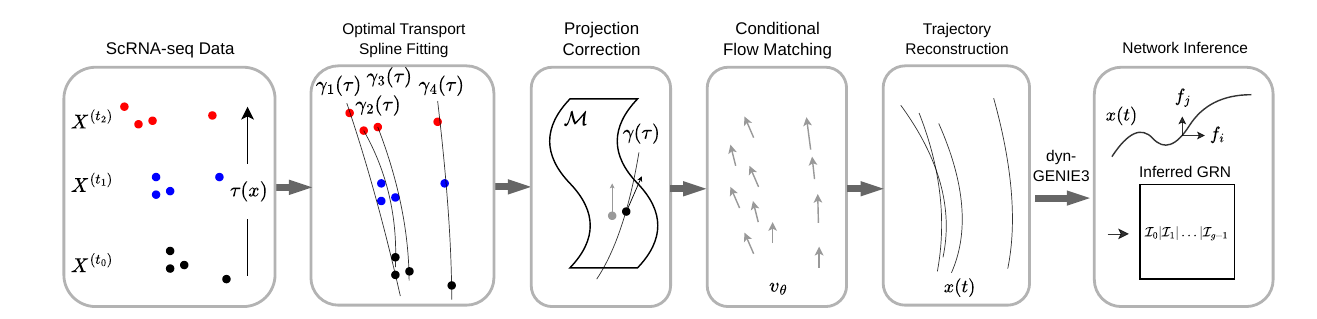}
  \caption{Graphical overview of FlowGRN+. 
           Cells are binned along pseudotime, coupled by OT, fitted with splines, refined by manifold projection, and then used to train a CFM model for trajectory reconstruction and downstream GRN inference.}
  \label{fig:model}
\end{figure*}

\subsection{Revisiting FlowGRN}
\label{ssec:revisiting_flowgrn}
FlowGRN~\cite{tong2025flowgrn} is the first GRN inference model that reconstructs cell trajectories using CFM and infers the GRN using dynGENIE3.
It is designed to capture key properties of scRNA-seq measurements for CFM training, such as dropouts and positive gene expression values.

Because gene expression measurements are corrupted by dropout and zero entries are unreliable for OT coupling, FlowGRN defines a dropout-robust cell similarity measure $d(x,y)$ that ignores dimensions with zero values.

Besides, since the vector field is generally defined on $\mathbb{R}^g$, whereas gene expression is non-negative, FlowGRN scales the expression values $x\in \mathbb{R}^g_{\geq 0}$ to a log domain $x'\in \mathbb{R}^g$ through a bijective, differentiable transformation $\mathcal{T}$:
\begin{equation}
  \label{eq:log-like-transformation}
  x' = \mathcal{T}(x) = 
    \begin{cases}
      \log(x+\epsilon) + 1, & \text{if } x < 1 \\
      x+\epsilon, & \text{if } x \geq 1,
    \end{cases}
\end{equation}
where $\epsilon$ is a positive value to avoid an undefined logarithm.

Overall, 
FlowGRN first computes the OT coupling $\pi_i$ between consecutive time points $t_i,t_{i+1}$ using $d$ as the metric, then transforms each cell $x$ into the log domain via $\mathcal{T}$.
FlowGRN is then trained under the [SF]\textsuperscript{2}M framework (an OTCFM model regularized by score matching) and learns a unified vector field $v_\theta$ that pushes each cell population $X^{(t_i)}$ to $X^{(t_{i+1})}$ in each time interval $[t_i,t_{i+1}]$.
Finally, the cell trajectories $x(t)$ are reconstructed by solving an initial value problem (IVP) starting from each $x$, followed by the inverse transformation $\mathcal{T}^{-1}$, and the GRN is inferred by applying dynGENIE3 to the reconstructed trajectories.

\subsection{Limitation of FlowGRN}
\label{ssec:limitation_flowgrn}
While FlowGRN shows top-tier performance on the BEELINE benchmark, it still faces limitations regarding the temporal coherence of the learned dynamics and human inspection.

\subsubsection{Dropout-sensitive references}
FlowGRN uses a single unified neural network to learn the vector field across all time intervals, but the reference velocity vectors are generated from linearly interpolated trajectories with nodes at the boundaries of the time intervals.
Thus, if a trajectory at one node is affected by dropouts, the reference velocity vectors will drastically change around the node, which introduces contradictory supervision signals for CFM training.

\subsubsection{Vector field discontinuity and loss of global dynamics}
Because linear interpolation only considers the pairwise local interval between two time points, the reference velocity vectors are discontinuous at the time interval boundaries.
This imposes no constraint on $v_\theta$ to learn a temporally coherent vector field across time intervals, because the vector field in each time interval is trained only to match local dynamics between corresponding time points, which is biologically unrealistic.
IMMFM~\cite{islam2025longitudinal} and 3MSBM~\cite{theodoropoulos2025momentum} also observe the same problem and mitigate it by applying pchip interpolation and modeling dynamics in the phase space, respectively.

\subsubsection{Reliance on clear snapshot distributions}
FlowGRN assumes that the cell distributions at each time point are well-separated, which is not always the case in real-world datasets, especially when the time points are close, and the developmental speed is non-uniform.
Thus, instead of using sampling time as a cell partition, FlowGRN relies on clustering algorithms to partition cells into ordered clusters for downstream OT coupling and CFM training.
This requires human inspection to select the best clustering parameters and results that align with biological understanding, making the process less reproducible and not automated.

\subsection{Our solution: FlowGRN+}
\label{ssec:solution_flowgrn+}
To overcome these limitations of FlowGRN, we build on the MMFM idea and propose FlowGRN+, an improved version of FlowGRN that integrates spline fitting and manifold projection into the CFM framework using pseudotime.
Spline fitting primarily improves the temporal coherence of the reference dynamics across pseudotime, whereas the projection scheme primarily improves manifold alignment by suppressing off-manifold components.
Our model, FlowGRN+, is trained with five steps in the log domain:
\begin{enumerate}
    \item Bins cells evenly by count along pseudotime and manually removes steady-state cells if the inferred pseudotime and sample time order do not match.
    \item Samples cells from different pseudotime bins with $z\sim q(z)$ and fits smoothing splines $\gamma(\tau)$ (see Section~\ref{ssec:spline_fitting}).
    \item Samples $\tau\sim\mathcal{U}(0,1)$ and constructs training pairs $x\sim p(\tau|z)$ and $u=u(x,\tau|z)$ using Eqs.~\eqref{eq:cfm_p}--\eqref{eq:cfm_u} with spline-defined $\mu(\tau|z)$ and $\sigma(\tau)$ from Eq.~\eqref{eq:mmfm_mu_sigma}.
    \item Projects $x$ and $u$ onto the data manifold and its tangent space, respectively, using the projection scheme detailed in Sections~\ref{ssec:projective_scheme} and~\ref{ssec:efficient_computation}.
    \item Trains the CFM model $v_\theta(x,\tau)$ to match the projected $u$ by minimizing the loss $\mathcal{L}_{MFM}(\theta)$ with regularization $\eta$ in Eq.~\eqref{eq:loss_mfm}.
\end{enumerate}

Similar to FlowGRN, the well-trained model $v_\theta$ will be used to reconstruct the cell trajectories by solving the IVP starting from each cell $x$ at $\tau(x)$:
\begin{equation}
  x(\tau+\Delta \tau) = x(\tau) + \int_\tau^{\tau+\Delta \tau} v_\theta(x(s),s) \, ds.
\end{equation}
In practice, trajectories are integrated using tangent-projected velocities, followed by retraction toward the data manifold at each numerical solver step.
The reconstructed trajectories are mapped back to the gene expression domain via $\mathcal{T}^{-1}$ and are used for GRN inference by dynGENIE3, as detailed in Section~\ref{ssec:dynGENIE3}.

\subsection{Spline fitting}
\label{ssec:spline_fitting}
We first bin the cells into $b$ bins evenly by count along the pseudotime $\tau(x)$, compute the OT coupling $\pi_i$ between consecutive bins, and sample cells $(x_0,\cdots,x_{b-1}) \sim q(z)=\prod_{i=0}^{b-2}\pi_i(x_i,x_{i+1})$.
Unlike FlowGRN, MMFM, or MMSFM, which use linear, cubic, and Hermite splines, respectively, we fit a smoothing spline on pseudotime $\gamma(\tau)$ to the samples $\{(\tau(x_i), x_i)\}_{i=0}^{b-1}$.
This spline is implemented as a B-spline with a regularization term $\lambda_{reg}$ that balances data fitting and spline curvature.
It minimizes the following objective:
\begin{equation}
    \sum_{i=0}^{b-1} \|x_i - \gamma(\tau(x_i))\|^2 + \lambda_{reg} \int \|\gamma''(\tau)\|^2 d\tau.
\end{equation}
This brings three advantages that solve the limitations of FlowGRN:
\begin{itemize}
    \item Each spline is fitted on a different set of pseudotime $\{\tau(x_i)\}_{i=0}^{b-1}$, so any discontinuity is scattered across the pseudotime range, reducing its impact on CFM training.
    \item Splines can link nonlocal observations across pseudotime, while smoothing splines can mitigate dropouts via curvature regularization, thus stabilizing the training signal.
    \item Our binning strategy is a stratified sampling strategy that does not rely on any clustering algorithm and is therefore more automated and reproducible.
\end{itemize}

However, spline fitting, especially cubic splines, may suffer from overshooting, which produces reference velocity vectors with substantial off-manifold components, thereby degrading CFM training.
Moreover, when integrating the learned vector field, the reconstructed trajectories may drift away from the support of the observed data distribution. % leading to inaccurate GRN inference.

\subsection{Projection scheme for reference velocity refinement}
\label{ssec:projective_scheme}

To alleviate this problem, we draw inspiration from weighted linear local tangent space alignment~\cite{shah2022weighted} and introduce a projection scheme to refine the reference velocity vectors by projecting them onto the tangent space $T\mathcal{M}$ of the data manifold $\mathcal{M}$.
This encourages CFM to learn a vector field better aligned with local manifold structure and to reconstruct trajectories that remain close to the data manifold.
Overall, to project a vector $v$ at a point $x$ onto $T\mathcal{M}$, we:
\begin{enumerate}
    \item use kNN neighbors $\mathcal{N}_k(x)$ to capture the local geometry of $\mathcal{M}$ at $x$,
    \item compute the weighted local covariance matrix $\Sigma(\mathcal{N}_k(x))$ as the local geometry descriptor,
    \item apply spectral filtering to suppress the eigenvectors in the normal direction and obtain the approximated projection matrix $P$, and
    \item construct an affine retraction operator $\Pi^R$ that maps off-manifold points back toward the data manifold.
\end{enumerate}

Specifically, to project a vector $v$ at a point $x$ onto $T\mathcal{M}$, we first find the kNN neighborhood $\mathcal{N}_k(x)$ of $x$ from all observed cells as the local patch of $\mathcal{M}$.
Then, we compute the local covariance matrix $\Sigma(x)$ weighted by an RBF kernel $w(x,y)=\exp(-\frac{\|x-y\|^2}{h(x)^2})$ for $y\in \mathcal{N}_k(x)$, where the bandwidth $h(x)$ is set as the median distance between $x$ and its neighbors.
Under a local linear approximation of the manifold, the dominant directions of variation align with the tangent space $T\mathcal{M}$, and thus the basis of $T\mathcal{M}$ can be approximated by the top-$m$ eigenvectors of $\Sigma(\mathcal{N}_k(x))$, where $m$ is the intrinsic dimension of $\mathcal M$.

However, the intrinsic dimension $m$ is unknown, and threshold-based eigenvalue selection, such as the explained variance ratio, is not robust and can induce jumps in the projection matrix across different points.
Therefore, we apply spectral filtering with shrinkage $g_{\lambda_0}(\lambda)=\frac{\lambda}{\lambda+\lambda_0}$ on the eigenvalues of $\Sigma(x)$, where $\lambda_0$ is a hyperparameter that controls the shrinkage strength.
If $\lambda\gg \lambda_0$, then $g_{\lambda_0}(\lambda)\approx 1$ and the corresponding eigenvectors are preserved, and vice versa.
Assuming the eigen-decomposition of $\Sigma$ is $U\Lambda U^T$, where $U$ is the matrix of eigenvectors and $\Lambda$ is the diagonal matrix of eigenvalues, we can compute the approximated projection matrix as $P(x)=Ug_{\lambda_0}(\Lambda)U^T$.
Equivalently,
\begin{equation}
P(x)=\Sigma\bigl(\Sigma+\lambda_0 I\bigr)^{-1}
      =I-\lambda_0\bigl(\Sigma+\lambda_0 I\bigr)^{-1}.
\end{equation}
Note that $P(x)$ is just an approximation of the projection matrix, which is not idempotent due to the shrinkage.

Similarly, to retract a point $x$ toward $\mathcal{M}$, we define an affine operator $\Pi$, and denote by $\Pi^R$ its $R$-fold composition:
\begin{equation}
    \Pi(x)=\overline{\mathcal{N}_k(x)}+P(x)(x-\overline{\mathcal{N}_k(x)}),
\end{equation}
where $\overline{\mathcal{N}_k(x)}$ is the mean of the neighbors of $x$.
This iterative retraction can be interpreted as a local curvature correction to the first-order tangent-space approximation so that the resulting point remains close to $\mathcal{M}$.

\begin{figure}[!t]
  \centering
  \includegraphics[width=0.95\linewidth,trim=7cm 8cm 4.5cm 5cm,clip]{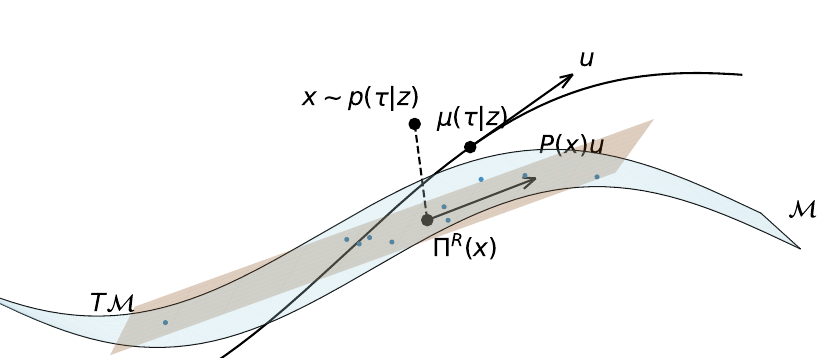}
  \caption{Illustration of the notation used in the projection scheme}
  \label{fig:manifold_projection}
\end{figure}

The above projection scheme, as illustrated in Fig.~\ref{fig:manifold_projection}, provides a practical link between the ambient space $\mathbb{R}^g$ and the data manifold $\mathcal{M}$.
However, the CFM loss in Eq.~\eqref{eq:loss_cfm} is defined in the ambient space, not in the tangent space.
We use the loss objective from MFM~\cite{kapusniak2024metric} with metric $G(x)=P(x)$ by taking expectation on $\tau\sim\mathcal{U}(0,1), z\sim q(z), x\sim p(\tau|z)$:
\begin{align}
    \mathcal{L}_{MFM}(\theta) = \mathbb{E} \| v_\theta(x,\tau) - u(x,\tau|z) \|^2_{G(x)}. \label{eq:loss_mfm}
\end{align}
However, this objective penalizes only the tangent-space component of $v_\theta$ and ignores the normal component, which may still drive trajectories away from $\mathcal{M}$ during integration, even though each numerical step is retracted toward $\mathcal{M}$.
Thus, we use a metric with a small regularization term $G(x)=P(x)+\eta I$.

\subsection{Efficient computation via the Woodbury identity}
\label{ssec:efficient_computation}
Directly computing $P(x)=\Sigma(\mathcal{N}_k(x))\bigl(\Sigma(\mathcal{N}_k(x))+\lambda_0 I\bigr)^{-1}$ requires solving a linear system in the ambient space $\mathbb{R}^g$, which is computationally expensive when $g$ is large.
However, since the weighted local covariance is constructed from only $k$ neighbors, it is intrinsically low-rank.

Let $\bar y_1,\dots,\bar y_k$ denote the demeaned neighbors of $x$, and let $\bar w_i(x)=\frac{w(x,y_i)}{\sum_{j=1}^k w(x,y_j)}$ be the normalized weights.
We define the weighted centered neighborhood matrix
\begin{equation}
Z(x)=
\begin{bmatrix}
\sqrt{\bar w_1(x)}\,\bar y_1^\top\\
\vdots\\
\sqrt{\bar w_k(x)}\,\bar y_k^\top
\end{bmatrix}
\in\mathbb{R}^{k\times g},
\end{equation}
so that $\Sigma(\mathcal{N}_k(x))=Z(x)^\top Z(x)$.
Using the Woodbury identity, $P(x)$ can be rewritten as
\begin{equation}
P(x)=
\frac{1}{\lambda_0}
Z(x)^\top
\left(
I_k+\frac{1}{\lambda_0}Z(x)Z(x)^\top
\right)^{-1}
Z(x).
\end{equation}
Therefore, instead of solving a $g\times g$ linear system, we only need to solve a $k\times k$ system, reducing the computational cost from $O(g^3)$ to $O(gk^2+k^3)$.
This is particularly beneficial in our setting, where the neighborhood size $k$ is much smaller than the ambient dimension $g$.

Furthermore, since the projection is computed at each step of the integration, a large portion of the computational cost comes from repeated kNN queries, which rarely change during integration when the change in $x$ is infinitesimal.
To optimize integration, we cache the $(k+1)$-NN neighbors and their distances $\{dist_i\}_{i=1}^{k+1}$ from the last query, and reuse the neighborhood for up to 100 iterations or until the change in $x$ exceeds the threshold $(dist_{k+1}-dist_k)/4$.

\begin{algorithm}[t]
\caption{Training and inference pipeline of FlowGRN+}
\label{alg:flowgrn_plus}
\small
\KwIn{Single-cell snapshots $X$ and pseudotime $\tau(x)$}
\KwOut{Reconstructed trajectories and inferred GRN}
\Repeat{convergence}{
    Sample cells from different pseudotime bins $z\sim q(z)$ and fit smoothing-spline trajectories $\gamma(\tau)$ as described in Section~\ref{ssec:spline_fitting}\;
    Sample $\tau\sim\mathcal{U}(0,1)$\;
    Compute $\mu(\tau|z)$ and $\sigma(\tau)$ with \eqref{eq:mmfm_mu_sigma}\;
    Sample $x\sim p(\tau|z)$ and reference velocity $u=u(x,\tau|z)$ according to \eqref{eq:cfm_p}--\eqref{eq:cfm_u}\;
    Project $x$ and $u$ onto the data manifold $\mathcal{M}$ and tangent space $T\mathcal{M}$ using the projective scheme in Sections~\ref{ssec:projective_scheme} and~\ref{ssec:efficient_computation}\;
    Update the CFM model $v_\theta(x,\tau)$ by minimizing the metric-aware loss $\mathcal L_{MFM}(\theta)$ in \eqref{eq:loss_mfm}\;
}
\ForEach{cell $x$ with pseudotime $\tau(x)$}{
    Reconstruct the trajectory by integrating the IVP\;
    $x(\tau+\Delta\tau)\gets x(\tau)+\int_{\tau}^{\tau+\Delta\tau} P(x(s)) v_\theta(x(s),s)\,ds$\;
}
Map reconstructed trajectories back to the gene expression domain via $\mathcal{T}^{-1}$\;
Infer the GRN from reconstructed trajectories using dynGENIE3 as described in Section~\ref{ssec:dynGENIE3}\;
\end{algorithm}

\section{Experiments}
\label{sec:experiments}
\subsection{Datasets and evaluation metrics}
\label{ssec:datasets}

We tested our method on experimental datasets in the BEELINE~\cite{pratapa2020benchmarking} benchmark, which are real gene expression profiles from sequencing data.
There are 7 sets of gene expression profiles from human and mouse cells:  hESC, hHep, mDC, mESC, mHSC-E, mHSC-GM, and mHSC-L.
The top 500 (1,000) high-variance genes and their corresponding transcription factors are selected from each dataset, matching the ``TFs + 500 (1,000) genes'' in the BEELINE benchmark.
Although BEELINE provides cell-type-specific and non-cell-type-specific GRNs as reference networks, they have not been updated since 2020.
We retrieved the gene regulations from the DoRothEA~\cite{garcia2019dorothea} and CollecTRI~\cite{muller2023collectri} databases and used them as reference networks.
Table~\ref{tab:experimental-network-statistics} describes their data statistics.

\begin{table}[!t]
\centering
\caption{Network Statistics of the Experimental Datasets}
\label{tab:experimental-network-statistics}
\begin{tabular}{lrrrr}
\toprule[1pt]\midrule[0.3pt]
Dataset      & \# Cells & \# Genes & \# Edges & \begin{tabular}[c]{@{}@{}r@{}}Network\\Density\\ ($\times 10^{-3}$)\end{tabular} \\
\hline
\multicolumn{5}{l}{\textbf{TFs + 500 genes}} \\
\quad hESC     & 758      & 1283     & 20007    & 12.15                                                                      \\
\quad hHep     & 425      & 1209     & 16915    & 11.57                                                                      \\
\quad mDC      & 383      & 1113     & 2854     & 2.30                                                                       \\
\quad mESC     & 421      & 501      & 139      & 0.55                                                                       \\
\quad mHSC-E   & 1071     & 763      & 945      & 1.62                                                                       \\
\quad mHSC-GM  & 889      & 698      & 676      & 1.39                                                                       \\
\quad mHSC-L   & 847      & 624      & 442      & 1.14                                                                       \\
\hline
\multicolumn{5}{l}{\textbf{TFs + 1,000 genes}} \\
\quad hESC    & 758      & 1783     & 26553    & 8.35                                                                       \\
\quad hHep    & 425      & 1709     & 22007    & 7.53                                                                       \\
\quad mDC     & 383      & 1613     & 3404     & 1.31                                                                       \\
\quad mESC    & 421      & 1001     & 416      & 0.42                                                                       \\
\quad mHSC-E  & 1071     & 1263     & 1135     & 0.71                                                                       \\
\quad mHSC-GM & 889      & 1198     & 895      & 0.62                                                                       \\
\quad mHSC-L  & 847      & 1124     & 729      & 0.58                                                                       \\
\midrule[0.3pt]\bottomrule[1pt]
\end{tabular}
\end{table}

We compare FlowGRN and FlowGRN+ with the baselines on the BEELINE benchmark, which includes LEAP~\cite{specht2017leap}, SCODE~\cite{matsumoto2017scode}, GRISLI~\cite{aubin2020grisli}, GRNVBEM~\cite{sanchez2018grnvbem}, SINCERITIES~\cite{papili2018sincerities}, Scribe~\cite{qiu2020scribe}, GENIE3~\cite{huynh2010genie3}, and GRNBOOST2~\cite{moerman2019grnboost2}.
Here, we excluded SCNS~\cite{woodhouse2018scns} as it requires prior knowledge of GRN.
Furthermore, we excluded ppcor~\cite{kim2015ppcor} and PIDC~\cite{chan2017pidc} because they cannot determine the direction of regulations, which is inconsistent with our research problem.

We evaluate the performance of FlowGRN+ and the baselines on the BEELINE benchmark using Area under Precision-
Recall Curve (AUPRC) and Early Precision Ratio (EPR).
The EPR is defined as the precision among the top $k$ edges normalized by the random precision, which emphasizes the precision of the inferred GRN and minimizes the false positive rate, with higher values indicating better performance and random performance corresponding to an EPR of 1.
AUPRC summarizes the precision-recall trade-off on all edges, whereas EPR focuses on the precision of top-ranked regulatory edges.

Besides, we also compare the smoothness of the reconstructed trajectories by computing the normalized total variation (TV) of the trajectories averaged across gene dimensions and time steps, which is defined as:
\[
\text{TV}(x) = \frac{1}{N\times g} \sum_{i=1}^{N-1} \|x(t_{i+1})-x(t_i)\|,
\]
for a trajectory with $N$ time points and $g$ genes.
However, we note that a smoother trajectory does not necessarily imply higher reconstruction fidelity.

\begin{table*}[!t]
\centering
\caption{AUPRC ($\times 10^{-3}$) of directed GRN inference models on different experimental datasets out of 5 runs.
         The \textbf{\underline{best}} and \underline{second-best} results are highlighted.
         ``OT'' and ``NE'' mean ``over time'' and ``numerical error'', respectively.
         }
\label{tab:AUPR-real}
\begin{tabular}{lrrrrrrr}
\toprule[1pt]\midrule[0.3pt]
Dataset & hESC & hHep & mDC & mESC & mHSC-E & mHSC-GM & mHSC-L \\
\midrule
\multicolumn{8}{l}{\textbf{TFs + 500 genes}} \\
 		\quad LEAP 		    & 11.91$\pm$0.00 		& \textbf{\underline{12.99$\pm$0.00}} 		& 2.38$\pm$0.00 		& \textbf{\underline{5.36$\pm$0.00}} 		& 1.94$\pm$0.00 		& 2.00$\pm$0.00 		& 1.28$\pm$0.00 \\
 		\quad SCODE 		  & 10.01$\pm$0.07 		& 10.57$\pm$0.05 		& 2.01$\pm$0.06 		& 0.85$\pm$0.02 		& 1.51$\pm$0.02 		& 1.30$\pm$0.05 		& 0.95$\pm$0.01 \\
 		\quad GRISLI 		  & 11.05$\pm$0.00 		& 10.57$\pm$0.00 		& \textbf{\underline{2.95$\pm$0.00}} 		& 0.51$\pm$0.00 		& 1.65$\pm$0.00 		& 1.21$\pm$0.00 		& 1.29$\pm$0.00 \\
    \quad GRNVBEM 		& OT             		& OT             		& 2.34$\pm$0.00 		& 0.55$\pm$0.00 		& 1.62$\pm$0.00 		& 1.39$\pm$0.00 		& 1.14$\pm$0.00 \\
 		\quad SINCERITIES & \underline{12.47$\pm$0.00} 		& 10.60$\pm$0.00 		& \underline{2.69$\pm$0.00} 		& 0.51$\pm$0.00 		& 1.49$\pm$0.00 		& 1.37$\pm$0.00 		& 1.18$\pm$0.00 \\
 		\quad Scribe 		  & 10.67$\pm$0.00 		& 9.64$\pm$0.00 		& 1.83$\pm$0.00 		& 0.53$\pm$0.00 		& 1.65$\pm$0.00 		& 1.38$\pm$0.00 		& 1.12$\pm$0.00 \\
		\quad GENIE3 		  & 10.50$\pm$0.01 		& 9.76$\pm$0.02 		& 2.10$\pm$0.01 		& \underline{3.02$\pm$0.12} 		& 2.94$\pm$0.48 		& 2.46$\pm$0.09 		& 1.93$\pm$0.09 \\
 		\quad GRNBOOST2 	& 11.96$\pm$0.02 		& 11.42$\pm$0.02 		& 2.32$\pm$0.02 		& 2.61$\pm$0.82 		& 1.79$\pm$0.03 		& 1.94$\pm$0.12 		& 1.45$\pm$0.16 \\
 		\quad FlowGRN     & \textbf{\underline{12.56$\pm$0.15}} 		& \underline{12.69$\pm$0.20} 		& 2.31$\pm$0.03 		& 1.88$\pm$0.18 		& \underline{3.43$\pm$0.28} 		& \textbf{\underline{3.45$\pm$0.09}} 		& \textbf{\underline{2.38$\pm$0.13}} \\
    \quad FlowGRN+ (Ours)     & 11.31$\pm$0.47 		& 12.62$\pm$0.17 		& 2.29$\pm$0.10 		& 1.82$\pm$0.17 		& \textbf{\underline{4.44$\pm$0.08}} 		& \underline{3.44$\pm$0.14} 		& \underline{1.95$\pm$0.08} \\
\midrule
\multicolumn{8}{l}{\textbf{TFs + 1,000 genes}} \\
   	\quad LEAP 		    & \underline{8.79$\pm$0.00} 		& \textbf{\underline{8.99$\pm$0.00}} 		& 1.38$\pm$0.00 		& \textbf{\underline{1.89$\pm$0.00}} 		& 1.01$\pm$0.00 		& 1.01$\pm$0.00 		& 0.66$\pm$0.00 \\
 		\quad SCODE 		  & 6.66$\pm$0.07 		& 6.76$\pm$0.10 		& 1.08$\pm$0.02 		& 0.52$\pm$0.01 		& 0.63$\pm$0.01 		& 0.58$\pm$0.01 		& 0.48$\pm$0.01 \\
 		\quad GRISLI 		  & 7.71$\pm$0.00 		& 7.01$\pm$0.00 		& \textbf{\underline{1.72$\pm$0.00}} 		& 0.42$\pm$0.00 		& 0.72$\pm$0.00 		& 0.55$\pm$0.00 		& 0.63$\pm$0.00 \\
    \quad GRNVBEM 		& OT             		& OT             		& OT             		& 0.42$\pm$0.00 		& OT             		& OT             		& 0.58$\pm$0.00 \\
 		\quad SINCERITIES & 8.57$\pm$0.00 		& 6.95$\pm$0.00 		& \underline{1.47$\pm$0.00} 		& 0.37$\pm$0.00 		& NE 		& 0.62$\pm$0.00 		& NE \\
 		\quad Scribe 		  & 7.30$\pm$0.00 		& 6.17$\pm$0.00 		& 0.99$\pm$0.00 		& 0.48$\pm$0.00 		& 0.71$\pm$0.00 		& 0.65$\pm$0.00 		& 0.63$\pm$0.00 \\
		\quad GENIE3 		  & 7.20$\pm$0.01 		& 6.25$\pm$0.01 		& 1.17$\pm$0.01 		& \underline{1.11$\pm$0.04} 		& 1.29$\pm$0.09 		& 1.13$\pm$0.02 		& 0.87$\pm$0.01 \\
 		\quad GRNBOOST2   & 8.29$\pm$0.01 		& 7.48$\pm$0.01 		& 1.33$\pm$0.01 		& 0.88$\pm$0.06 		& 0.81$\pm$0.01 		& 0.86$\pm$0.05 		& 0.69$\pm$0.04 \\
 		\quad FlowGRN	    & \textbf{\underline{8.88$\pm$0.17}} 		& \underline{8.69$\pm$0.56} 		& 1.28$\pm$0.03 		& 0.82$\pm$0.05 		& \textbf{\underline{1.62$\pm$0.15}} 		& \textbf{\underline{1.60$\pm$0.06}} 		& \underline{0.91$\pm$0.03} \\
    \quad FlowGRN+ (Ours)	    & 7.87$\pm$0.20 		& 8.30$\pm$0.15 		& 1.46$\pm$0.08 		& 0.93$\pm$0.03 		& \underline{1.57$\pm$0.05} 		& \underline{1.51$\pm$0.07} 		& \textbf{\underline{0.95$\pm$0.03}} \\
\midrule[0.3pt]\bottomrule[1pt]
\end{tabular}
\end{table*}

\begin{table*}[!ht]
\centering
\caption{EPR of directed GRN inference models on different experimental datasets out of 5 runs with different random seeds.
         The \textbf{\underline{best}} and \underline{second-best} results are highlighted.
         ``OT'' and ``NE'' mean ``over time'' and ``numerical error'', respectively.}
\label{tab:EPR-real}
\begin{tabular}{lrrrrrrr}
\toprule[1pt]\midrule[0.3pt]
Dataset & hESC & hHep & mDC & mESC & mHSC-E & mHSC-GM & mHSC-L \\
\midrule
\multicolumn{8}{l}{\textbf{TFs + 500 genes}} \\
 		\quad LEAP 		    & 0.94$\pm$0.00 		& \textbf{\underline{1.84$\pm$0.00}} 		& 1.98$\pm$0.00 		& \textbf{\underline{38.97$\pm$0.00}} 		& \textbf{\underline{5.87$\pm$0.00}} 		& 3.20$\pm$0.00 		& 1.99$\pm$0.00 \\
 		\quad SCODE 		  & 0.39$\pm$0.06 		& 0.97$\pm$0.06 		& 0.00$\pm$0.00 		& 0.00$\pm$0.00 		& 0.00$\pm$0.00 		& 2.13$\pm$1.27 		& 0.00$\pm$0.00 \\
 		\quad GRISLI 		  & 0.94$\pm$0.00 		& 0.91$\pm$0.00 		& \underline{2.13$\pm$0.00} 		& 0.00$\pm$0.00 		& 1.96$\pm$0.00 		& 1.07$\pm$0.00 		& 5.98$\pm$0.00 \\
    \quad GRNVBEM 		& OT             		& OT 		            & 0.76$\pm$0.00 		& 0.00$\pm$0.00 		& 1.96$\pm$0.00 		& 1.07$\pm$0.00 		& 0.00$\pm$0.00 \\
 		\quad SINCERITIES & \textbf{\underline{1.34$\pm$0.00}} 		& 0.51$\pm$0.00 		& 1.06$\pm$0.00 		& 0.00$\pm$0.00 		& 0.65$\pm$0.00 		& 0.00$\pm$0.00 		& 0.00$\pm$0.00 \\
 		\quad Scribe 		  & 0.67$\pm$0.00 		& 0.61$\pm$0.00 		& 0.00$\pm$0.00 		& 0.00$\pm$0.00 		& 1.30$\pm$0.00 		& 1.07$\pm$0.00 		& 0.00$\pm$0.00 \\
		\quad GENIE3 		  & 0.82$\pm$0.01 		& 1.06$\pm$0.02 		& 1.98$\pm$0.24 		& 5.20$\pm$7.12 		& 5.35$\pm$0.29 		& 7.89$\pm$0.58 		& 5.98$\pm$0.00 \\
 		\quad GRNBOOST2   & 0.78$\pm$0.02 		& 0.96$\pm$0.02 		& 1.43$\pm$0.17 		& \underline{15.59$\pm$16.94} 	& 2.87$\pm$0.74 		& 8.10$\pm$0.58 		& \underline{6.78$\pm$1.78} \\
 		\quad FlowGRN     & 1.14$\pm$0.12 		& 1.36$\pm$0.12 		& 0.64$\pm$0.13 		& 0.00$\pm$0.00 		& 5.61$\pm$0.58 		& \underline{9.60$\pm$0.75} 		& \textbf{\underline{7.18$\pm$1.78}} \\
    \quad FlowGRN+ (Ours)     & \underline{1.18$\pm$0.16} 		& \underline{1.84$\pm$0.08} 		& \textbf{\underline{2.16$\pm$0.39}} 		& 2.60$\pm$5.81 		& \underline{5.74$\pm$0.85} 		& \textbf{\underline{10.87$\pm$0.89}} 		& 6.38$\pm$0.89 \\
\midrule
\multicolumn{8}{l}{\textbf{TFs + 1,000 genes}} \\
 		\quad LEAP 		    & 1.28$\pm$0.00 		& \textbf{\underline{2.36$\pm$0.00}} 		& 2.92$\pm$0.00 		& \textbf{\underline{46.32$\pm$0.00}} 		& 8.67$\pm$0.00 		& 5.38$\pm$0.00 		& 0.00$\pm$0.00 \\
 		\quad SCODE 		  & 0.29$\pm$0.06 		& 1.03$\pm$0.06 		& 0.00$\pm$0.00 		& 0.00$\pm$0.00 		& 0.00$\pm$0.00 		& 3.58$\pm$1.27 		& 0.00$\pm$0.00 \\
 		\quad GRISLI 		  & 0.82$\pm$0.00 		& 0.93$\pm$0.00 		& 2.92$\pm$0.00 		& 0.00$\pm$0.00 		& 1.24$\pm$0.00 		& 0.00$\pm$0.00 		& 2.38$\pm$0.00 \\
    \quad GRNVBEM 		& OT             		& OT             		& OT             		& 0.00$\pm$0.00 		& OT             		& OT             		& 0.00$\pm$0.00 \\
 		\quad SINCERITIES & 1.42$\pm$0.00		& 0.58$\pm$0.00 		& 0.22$\pm$0.00 		& 0.00$\pm$0.00 		& NE 		  & 0.00$\pm$0.00 		& NE \\
 		\quad Scribe 		  & 0.69$\pm$0.00 		& 0.60$\pm$0.00 		& 0.22$\pm$0.00 		& \underline{11.58$\pm$0.00} 		& 2.48$\pm$0.00 		& 0.00$\pm$0.00 		& 0.00$\pm$0.00 \\
		\quad GENIE3 		  & 1.10$\pm$0.02 		& 1.30$\pm$0.04 		& 2.87$\pm$0.33 		& 6.95$\pm$4.84 		& 8.92$\pm$1.04 		& 13.26$\pm$0.98 		& 7.13$\pm$0.00 \\
 		\quad GRNBOOST2   & 0.95$\pm$0.04 		& 1.09$\pm$0.04 		& 1.80$\pm$0.42 		& 10.42$\pm$4.84 		& 6.44$\pm$1.61 		& 11.11$\pm$1.96 		& \textbf{\underline{8.08$\pm$1.30}} \\
 		\quad FlowGRN	    & \textbf{\underline{1.59$\pm$0.22}} 		& 2.09$\pm$0.38 		& 0.85$\pm$0.37 		& 1.16$\pm$2.59 		& \underline{11.39$\pm$0.55} 		& \textbf{\underline{16.13$\pm$0.00}} 		& \underline{7.61$\pm$1.06} \\
    \quad FlowGRN+ (Ours)	    & \underline{1.55$\pm$0.21} 		& \underline{2.29$\pm$0.11} 		& \textbf{\underline{3.28$\pm$1.10}} 		& 8.11$\pm$3.17 		& \textbf{\underline{12.38$\pm$0.00}} 		& \underline{16.13$\pm$1.27} 		& 7.13$\pm$0.00 \\
\midrule[0.3pt]\bottomrule[1pt]
\end{tabular}
\end{table*}

\subsection{Experimental setup}
\label{ssec:experimental-setup}
We use Slingshot~\cite{street2018slingshot} to compute pseudotime for all datasets, and normalize it to $[0,1]$ for CFM training.
During trajectory reconstruction, $v_\theta$ in the OTCFM model is implemented as a ResNet~\cite{he2016deep} with 20 hidden layers, each with 128 hidden units and SELU activations, optimized by AdamW with a learning rate of $10^{-4}$ decayed by a factor of 2 every 100 epochs.
The hyperparameters in FlowGRN+ are set as follows: $b=6$ for binning, $\sigma=0.01$, $\lambda_0=0.001$, $\eta=0.001$, $k=15$ for kNN, $R=2$ for retraction on $u$, and $R=3$ for retraction on $x$.
$\lambda_{reg}$ in smoothing spline fitting is chosen by generalized cross-validation.
Models are trained for 1,000 epochs with a batch size of 128 samples.
Trajectories are integrated using the DOPRI5 solver in TorchDyn~\cite{poli2020torchdyn}.
All neural networks are trained on an NVIDIA V100 SXM2 GPU with 16GB memory.

In GRN inference, dynGENIE3 is used with the default parameters: 1,000 random forest trees, each trained with $\sqrt{g}$ genes.
dynGENIE3 is trained on a computing node with 128 CPUs at 2.6GHz and 256GB of memory.
All baselines are also run on the same platform within a 48-hour interval, using their default parameters, wrapped in a Singularity container to ensure reproducibility.
Each method is run 5 times with random seeds 1--5 on each dataset, and the average and standard deviation of EPR are reported.

\section{Results and Discussion}
\label{sec:results_discussion}

\subsection{FlowGRN+ achieves top-tier performance on the BEELINE benchmark}

Tables~\ref{tab:AUPR-real}--\ref{tab:EPR-real} report the AUPRC and EPR of FlowGRN+, FlowGRN, and the baselines on the BEELINE experimental datasets.
Overall, FlowGRN+ achieves top-tier performance on the experimental datasets, ranking among the top two models on 6 out of 14 settings for AUPRC and 10 of the 14 settings for EPR.
Compared directly with FlowGRN, FlowGRN+ achieves a higher EPR in 10 of the 14 settings, whereas its AUPRC is higher in only 4 of the 14 settings.
This suggests that its downstream advantage is concentrated in recovering top-ranked edges and does not extend consistently across the full ranked edge list.

Notably, FlowGRN+ outperforms FlowGRN on mDC and mESC datasets.
We speculate that this improvement is partly due to the removal of stationary cells, because near equilibrium the dynamical signal is weak and pseudotime-based temporal ordering becomes unreliable.

\subsection{FlowGRN+ with projection recovers smoother trajectories}

\begin{figure*}
    \centering
    \includegraphics[width=\textwidth]{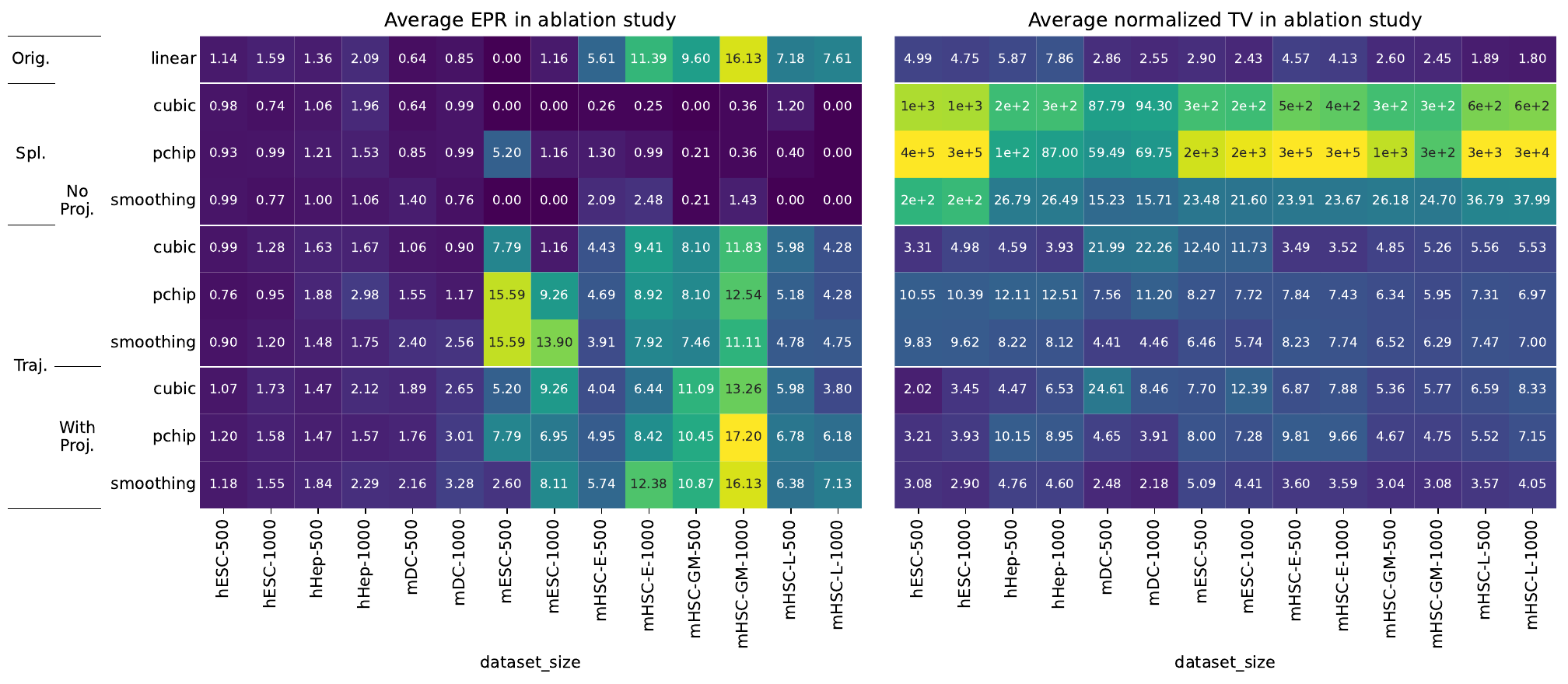}
    \caption{Ablation studies on the trajectories reconstructed by FlowGRN (Orig.), spline fitting (Spl.) and FlowGRN+ (Traj.), with or without projection scheme.
             Rows are grouped by model family and projection setting, and columns correspond to dataset-size combinations.
             EPR and normalized TV are averaged across 5 random seeds for each setting.
             Color scale in normalized TV is in log scale and capped at 3,000 for better visualization.}
    \label{fig:traj-tv}
\end{figure*}

In this work, we define the default FlowGRN+ setting as multi-marginal OT-chain sampling with smoothing splines and manifold projection, and treat remaining variants as ablations.

For trajectories reconstructed by FlowGRN (Orig.), spline fitting (Spl.), and FlowGRN+ (Traj.), we examine in Fig.~\ref{fig:traj-tv} how the projection scheme and the choice of spline fitting method affect the average EPR and normalized TV.
A higher EPR indicates more precise GRN inference, and a lower normalized TV indicates smoother trajectories.

\subsubsection{Spline fitting versus CFM models}
We observe that spline fitting without CFM learning has a nearly-zero EPR and a much higher normalized TV.
This is expected because spline fitting only interpolates each gene's expression trajectory as a function of pseudotime, whereas CFM-based models (FlowGRN and FlowGRN+) learn a coupled, state-dependent vector field that captures cross-gene dependencies.

\subsubsection{CFM models with and without projection}
Across 42 settings, projection yields higher EPR in 28 settings (66.67\%) and lower normalized TV in 30 settings (71.43\%), empirically supporting the view that the projection scheme often improves the learned dynamics and GRN inference performance.

\subsubsection{Spline fitting methods on projected CFM models}
In general, smoothing splines have a higher EPR average rank ($1.64$) across datasets than pchip ($2.00$) and cubic ($2.36$) splines, and a consistently lower normalized TV average rank ($1.14$) than cubic ($2.36$) and pchip ($2.50$) splines.
A plausible explanation is that smoothing splines do not need to pass through all data points and are therefore less prone to overfitting to dropout values and overshooting than interpolation-based methods.

\section{Conclusion and Limitations}
\label{sec:conclusion}
In this work, we have presented FlowGRN+, an improved framework for GRN inference from scRNA-seq data that incorporates spline fitting and manifold projection to improve the temporal coherence of the learned dynamics and the manifold alignment of reconstructed trajectories.
It achieves top-tier performance on the experimental BEELINE benchmark, and the projection scheme often improves the smoothness of the reconstructed trajectories.

Compared with FlowGRN, our new method reduces the reliance on manual clustering-based partitioning of cells into distinct distributions and is therefore more automated and reproducible.
In addition, the learned dynamics are more temporally coherent and give smoother trajectory reconstruction.

However, the projection scheme also has limitations.
First, the smooth manifold assumption may be violated in branching datasets, and the uniqueness of ODE solutions can further limit the modeling of branching trajectories, potentially contributing to the performance drop.
Second, the projection scheme is computationally expensive, resulting in a longer training time than FlowGRN.
A tighter integration of the projection scheme with the CFM model remains an important future direction.

In the future, we will explore how to better extract networks from the learned dynamics.
In our experiments, GRNBOOST2 yields lower AUPRC than GENIE3 in 8 of the 14 settings and lower EPR in 9 of the 14 settings, although both follow a per-target supervised-learning paradigm for GRN inference.
This suggests that the choice of model class and explanation method can materially affect downstream network recovery.
We will also explore how to constrain the CFM model to align with known biological pathways, so that we can leverage prior knowledge to improve GRN inference.

\section*{Conflicts of Interest}
The authors declare no conflicts of interest.

\section*{Data and Code Availability}
We do not generate any new data in this work. 
All the data used in this work are publicly available. 
The code of FlowGRN+ is implemented in Python 3.13 and is available at \url{https://github.com/1250326/FlowGRN_plus}.

% \section*{Acknowledgment}

% The preferred spelling of the word ``acknowledgment'' in America is without 
% an ``e'' after the ``g''. Avoid the stilted expression ``one of us (R. B. 
% G.) thanks $\ldots$''. Instead, try ``R. B. G. thanks$\ldots$''. Put sponsor 
% acknowledgments in the unnumbered footnote on the first page.

% \section*{References}
\bibliographystyle{IEEEtran}
\bibliography{references.bib}

\end{document}